\documentclass{ifacconf}

\usepackage{graphicx}      %
\usepackage{subcaption}
\usepackage[justification=centering]{caption}
\usepackage{natbib}        %
\usepackage{siunitx} %
\usepackage{comment} %
\usepackage{algorithm}
\usepackage[noend]{algpseudocode}
\usepackage{booktabs}
\usepackage{url}

\theoremstyle{definition}

\newtheorem{definition}{Definition}

\usepackage{mathtools}
\theoremstyle{plain}

\newcounter{problem}
\newenvironment{problem}{%
\par\vspace{3pt}\noindent\refstepcounter{problem}\textbf{Problem~\theproblem:}}%

\usepackage[extramath,probability,reviewmark]{mylatexdefs}

\newcommand{\CAV}[1]{CAV\textendash\ensuremath{#1}\xspace}

\newcommand{\lane}[1]{lane\textendash\ensuremath{#1}\xspace}

\newcommand{\bbsym}[1]{\ensuremath{\boldsymbol{#1}}}

\begin{document}
\begin{frontmatter}

\title{Accelerating Time-Optimal Trajectory Planning for Connected and Automated Vehicles with Graph Neural Networks
} 

\thanks[footnoteinfo]{This research was supported in part by NSF under Grants CNS-2149520, CMMI-2348381, IIS-2415478, and in part by Mathworks.}
\thanks[footnoteinfo]{This work was conducted while Viet-Anh Le was at Cornell University.}

\author[First]{Viet-Anh Le} and
\author[Second]{Andreas A. Malikopoulos} 

\address[First]{University of Pennsylvania, 
   Philadelphia, PA 19104 USA\\(e-mail: vietanh@seas.upenn.edu).}
\address[Second]{Cornell University, 
   Ithaca, NY 14850 USA\\(e-mail: amaliko@cornell.edu)}

\begin{abstract} %
In this paper, we present a learning-based framework that accelerates time- and energy-optimal trajectory planning for connected and automated vehicles (CAVs) using graph neural networks (GNNs). 
We formulate the multi-agent coordination problem encountered in traffic scenarios as a cooperative trajectory planning problem that minimizes travel time, subject to motion primitives derived from energy-optimal solutions. 
The performance of this framework can be further improved through replanning at each time step, enabling the system to incorporate newly observed information.
To achieve real-time execution, we employ a graph isomorphism network with edge features (GINEConv) to learn the solutions of the time-optimal trajectory planning problem from offline-generated data. 
The trained model produces online predictions that serve as warm-starts for numerical optimization, thereby enabling rapid computation of minimal exit times and the associated feasible trajectories. 
This learning-to-warm-start approach substantially reduces computation time while preserving the control performance of the time- and energy-optimal trajectory planning framework.

\end{abstract}


\end{frontmatter}

\section{Introduction}

The rapid advancements in vehicle connectivity and automation offer promising opportunities to enhance safety while reducing energy consumption, greenhouse gas emissions, and travel delays; see \citep{kopelias2020connected}. 
A growing body of research has highlighted the benefits of coordinating connected and autonomous vehicles (CAVs) through control and optimization techniques in a wide range of traffic scenarios.
Several effective approaches have been proposed, including hierarchical optimization for scheduling and planning; see \citep{chalaki2020TITS,xiao2019decentralized}, distributed model predictive control; see \citep{kloock2019distributed,katriniok2022fully}, and multi-agent reinforcement learning; see \citep{hua2025multi,zhang2023coordinating}.

Trajectory planning based on time- and energy-optimal control has also been widely applied across a range of traffic scenarios, both for fully automated traffic streams; see \citep{Malikopoulos2020}, and for mixed-traffic environments; see \citep{le2024stochastic,Le2023CDC}. 
The effectiveness of this trajectory planning framework can be further enhanced through \emph{replanning}, either at fixed periodic intervals or in response to specific events; see \citep{chalaki2021Reseq,le2024stochastic}. 
Replanning improves performance and introduces feedback into the control architecture, enhancing robustness to disturbances and modeling uncertainties. 
However, at each replanning instance, all CAVs must sequentially re-solve their time-optimal control problems using the latest vehicle states. 
As the number of CAVs grows, this sequential optimization becomes increasingly computationally demanding, limiting real-time deployment.

In this work, we develop a \emph{learning-to-warm-start} framework to accelerate finding the solutions of time and energy-optimal trajectory planning for CAVs based on graph neural networks (GNNs). 
In particular, we train a GINEConv network, a GNN architecture that incorporates edge attributes into the message-passing process, using offline data to learn the optimal solutions of the time-optimal cooperative trajectory planning problem for a group of CAVs.
In real-time implementation, the predicted optimal terminal time for each CAV can be used as an initial guess for a numerical algorithm to quickly find the optimal solution for the terminal time and the corresponding optimal trajectory that satisfies all the constraints.
The proposed framework enables fast implementation for solving the cooperative time-optimal trajectory planning problem, in which the problems for multiple CAVs must be solved sequentially.

The remainder of this paper is organized as follows. 
Section~\ref{sec:prb} presents the multi-agent time-optimal trajectory planning framework for CAV coordination at an unsignalized intersection.
Section~\ref{sec:gnn} develops a GNN-based method to learn the optimal time solutions, which can serve as warm-start solutions for the numerical algorithm.
Finally, we provide the simulation results in Section~\ref{sec:sim} and concluding remarks in Section~\ref{sec:cls}.

\section{Coordination of Connected and Automated Vehicles}
\label{sec:prb}

In this section, we describe the problem and summarize the time-optimal trajectory planning framework
to coordinate the CAVs, developed in our previous work.

\subsection{Problem Formulation}

\begin{figure}[tb]
\centering
\includegraphics[width=0.49\textwidth,trim=100 90 160 100, clip]{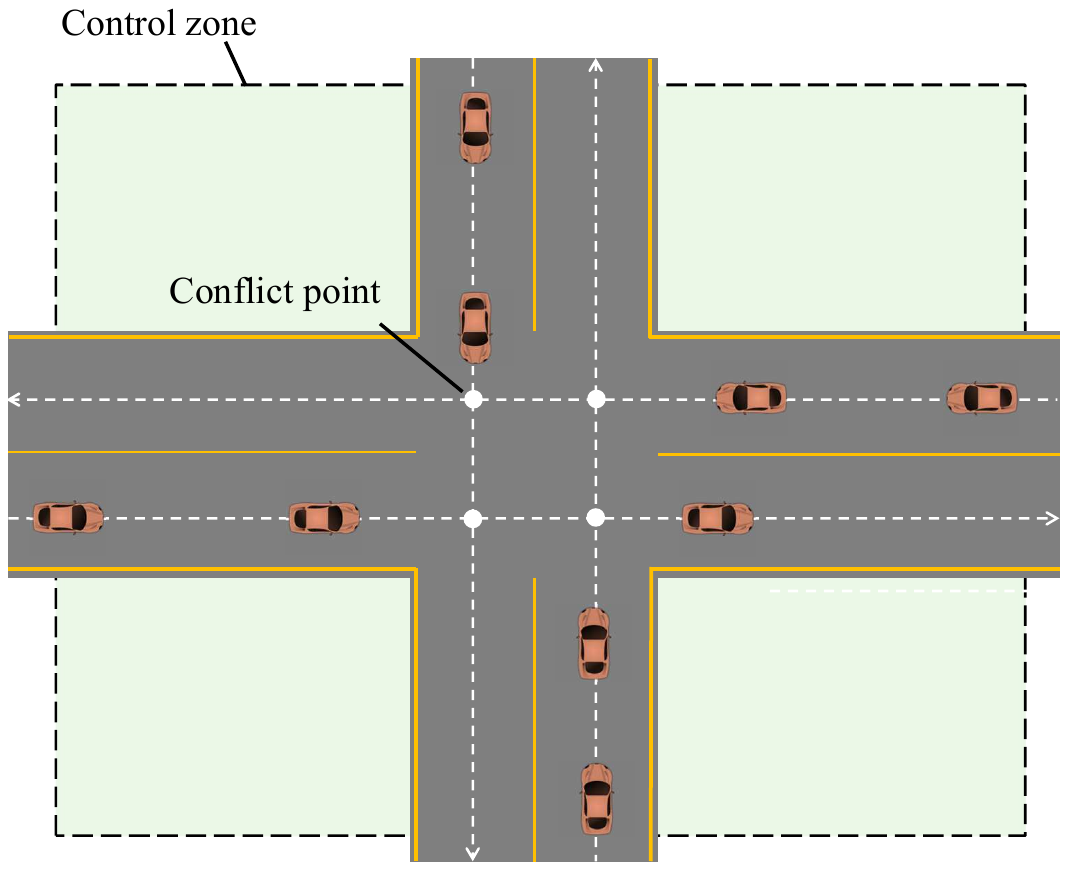}
\caption{An intersection scenario with 4 lanes.}
\label{fig:scenario}
\end{figure}

We consider the problem of coordinating multiple CAVs in a single-lane unsignalized intersection, as illustrated in Fig.~\ref{fig:scenario}. 
The points at which the paths of different CAVs intersect, where a lateral collision may occur, are referred to as \emph{conflict points}. 
Although an unsignalized intersection serves as the representative scenario in this work, the proposed framework can be readily extended to other environments featuring lateral conflicts, such as merging roadways or roundabouts.
We define a \emph{control zone} within which the CAVs operate under the proposed coordination framework. 
We assume that a centralized coordinator is available, with access to the positions of all CAVs within the control zone. 
Moreover, the CAVs and the coordinator are able to exchange information while they remain inside the control zone.

Next, we provide some necessary definitions for the exposition.
\begin{definition}
(Lanes) Let \lane{l} be the $l$-th lane in the scenario and $\LLL$ be the set of all lanes' indices.
We set each lane's origin location at the control zone's entry.
\end{definition}

\begin{definition}
(Conflict points) Let point $n$, and the notation $n=l\otimes m$ denote that point $n$ is the conflict point between \lane{l} and \lane{m}. 
Let $\psi_l^n$ and $\psi_m^n$ be the positions of point $n$ along \lane{l} and \lane{m}, respectively.
\end{definition}

\begin{definition}
(Vehicles)
\label{def:set_vehicle}
Let ${\AAA(t) = \{1,\ldots,N(t)\}}$, ${t\in\mathbb{R}_{\ge0}}$, be the set of CAVs traveling inside the control zone, where ${N(t)\in\mathbb{N}}$ is the total number of vehicles. 
Note that the indices of the vehicles are determined by the order in which they enter the control zone. 
\end{definition}

Let $p^0$ and $p^{f} \in \RR$ be the positions of the control zone entry and exit, respectively.
We consider that the dynamics of each vehicle ${i \in \AAA (t)}$ are described by a double integrator model as follows
\begin{equation}\label{eq:model2}
\begin{split}
\dot{p}_{i}(t) &= v_{i}(t), 
\\
\dot{v}_{i}(t) &= u_{i}(t), 
\end{split} 
\end{equation}
where ${p_{i}\in\mathcal{P}}$, ${v_{i}\in\mathcal{V}}$, and
${u_{i}\in\mathcal{U}}$ denotes the longitudinal position of the rear bumper, speed, and control input (acceleration/deceleration) of the vehicle, respectively. 
The sets ${\mathcal{P}, \mathcal{V},}$ and $\mathcal{U}$ are compact subsets of $\RR$. 
The control input is bounded by 
\begin{equation}\label{eq:uconstraint}
u_{\min} \leq u_{i}(t)\leq u_{\max},\quad \forall i \in \LLL(t),
\end{equation}
where ${u_{\min}<0}$ and ${u_{\max}>0}$ are the minimum and maximum control inputs, respectively, as designated by the physical acceleration and braking limits of the vehicles, or limits that can be imposed for driver/passenger comfort.
Next, we consider the speed limits of the CAVs, 
\begin{equation}\label{eq:vconstraint} 
v_{\min} \leq v_i(t) \leq v_{\max},\quad \forall i \in \LLL (t), 
\end{equation}
where ${v_{\min}> 0}$ and ${v_{\max}>0}$ are the minimum and maximum allowable speeds. 

Next, let $t_i^0$ and $t_i^{f} \in \mathbb{R}_{\ge0}$ be the times at which each vehicle $i$ enters and exits the control zone, respectively.
To avoid collisions between vehicles in the control zone, we impose two constraints: (1) lateral constraints between vehicles traveling on different lanes and (2) rear-end constraints between vehicles traveling on the same lane.
Specifically, to prevent a potential conflict between \CAV{i} and \CAV{k} traveling on \lane{l} and \lane{m} with a conflict point $n$, we require a minimum time gap ${\delta_{l} \in \mathbb{R}_{\ge 0}}$ between the time those CAVs cross the conflict point.
Let ${t_i^{n}}$ and $t_k^{n}$ be the time when the \CAV{i} and \CAV{k} cross the conflict point $n$.
Note that since $0 < v_{\min} \le v_i(t)$, the position $p_i(t)$ is a strictly increasing function. 
Thus, the inverse function $t_i(\cdot) = p_i^{-1} (\cdot)$ exists and there exist unique value of $t_i^{n}$ such that $p (t_i^{n}) = \psi_l^n$, and we impose the following lateral constraint, 
\begin{equation}
\label{eq:lateral_constraint}
\abs{t_i^{n} - t_k^{n}} \geq \delta_{l}. 
\end{equation}

Additionally, to prevent rear-end collision between \CAV{i} and its immediate preceding \CAV{k} traveling on the same lane, we impose the following rear-end safety constraint:
\begin{equation}
\label{eq:rearend_constraint}
 p_{k}(t - \delta_{r}) - p_{i}(t) \geq d_{\min},\, t \in [t_i^{0}, t_k^{f}],
\end{equation}
where ${d_{\min} \in \mathbb{R}_{\ge0}}$ and ${\delta_{r} \in \mathbb{R}_{\ge0}}$ are the minimum distance at a standstill and safe time gap. Note that $p_{k}(t - \delta_{r})$ denotes the position of \CAV{k} at time instant $t - \delta_{r}$. 

\subsection{Time-Optimal Trajectory Planning}

Next, we summarize the cooperative time-optimal trajectory planning framework developed for coordinating CAVs; see \citep{Malikopoulos2020}.
We start the exposition with the unconstrained solution of an energy-optimal control problem for each \CAV{i}; see \citep{malikopoulos2018decentralized}.
Given a fixed $t_i^{f}$ that \CAV{i} exits the control zone, the energy-optimal control problem aims at finding the optimal control input (acceleration/deceleration) for each CAV by solving the following problem.
\begin{problem}
\label{prb:energy-optimal-1}
(\textbf{Energy-optimal control problem})
Let $t^0$ be the current time and $t_{i}^{f}$ be the time that \CAV{i} exits the control zone. 
The energy-optimal control problem for \CAV{i} at $t$ is given by: 
\begin{equation}
\label{eq:energy_cost}
\begin{split}
&\minimize_{u_i(t)\in \UUU} \quad \frac{1}{2} \int_{t^0}^{t^{f}_{i}} u^2_i(t)~\mathrm{d}t,
\\ 
& \subjectto 
\\
& \quad \eqref{eq:model2}, \eqref{eq:uconstraint},\eqref{eq:vconstraint},
\eqref{eq:rearend_constraint}, 
\\
& \text{given:} 
\\
& \quad p_i (t_i^0) = p_i^0, v_i (t_i^0) = v_i^0, p_i (t_i^{f}) = p^{f},
\end{split}
\end{equation}
where $v_i^0$ is the current speed of \CAV{i}.
The boundary conditions in \eqref{eq:energy_cost} are set at the current and exit of the control zone. 
\end{problem}

The closed-form solution of Problem~\ref{prb:energy-optimal-1} for each \CAV{i} can be derived using the Hamiltonian analysis.
If none of the state and control constraints are active, the closed-form optimal control law and trajectory are given by; see \citep{Malikopoulos2020}
\begin{equation}\label{eq:optimalTrajectory}
\begin{split}
u_i(t) &= 6 \phi_{i,3} t + 2 \phi_{i,2}, \\
v_i(t) &= 3 \phi_{i,3} t^2 + 2 \phi_{i,2} t + \phi_{i,1}, 
\\
p_i(t) &= \phi_{i,3} t^3 + \phi_{i,2} t^2 + \phi_{i,1} t + \phi_{i,0},
\end{split}
\end{equation} 
where ${\phi_{i,3}, \phi_{i,2}, \phi_{i,1}, \phi_{i,0} \in \RR}$ are constants of integration.
Since the speed of \CAV{i} is not specified at the exit time $t_i^{f}$, we consider the boundary condition 
\begin{equation}
\label{eq:uif}
{u_i(t_i^{f}) = 0}.
\end{equation}
For the full derivation of the closed-form solution in \eqref{eq:optimalTrajectory} using Hamiltonian analysis, the readers are referred to \cite{Malikopoulos2020}.

Given the boundary conditions in \eqref{eq:energy_cost} and \eqref{eq:uif}, and considering $t_i^{f}$ is known, the constants of integration can be found by:
\begin{equation}
\label{eq:phi_i}
\bbsym{\phi}_i =
\begin{bmatrix}
\phi_{i,3}
\\
\phi_{i,2}
\\
\phi_{i,1}
\\
\phi_{i,0}
\end{bmatrix}
= 
\begin{bmatrix}
(t_i^0)^3 & (t_i^0)^2 & t_i^0 & 1 
\\
3(t_i^0)^2 & 2t_i^0 & 1 & 0 
\\
(t_i^{f})^3 & (t_i^{f})^2 & t_i^{f} & 1 
\\
6t_i^{f} & 2 & 0 & 0 
\end{bmatrix}^{-1}
\begin{bmatrix}
p^0 
\\
v_i^0 
\\
p^{f} 
\\
0
\end{bmatrix}.
\end{equation}

Note that, as the position is an increasing function of time, we can derive the inverse function that represents the time trajectory $t_i(p_i)$ as a function of the position. %

Next, we formulate the time-optimal control problem to minimize the travel time and guarantee all the constraints for CAVs given the energy-optimal trajectory \eqref{eq:optimalTrajectory} at $t_i^0$.
We enforce this unconstrained trajectory as a motion primitive to avoid the complexity of solving a constrained optimal control problem by piecing constrained and unconstrained arcs together; see \citep{malikopoulos2018decentralized}.

\begin{problem} 
\label{prb:2}
(\textbf{Time-optimal trajectory planning})
At the time $t_i^{0}$ of entering the control zone, let ${\mathcal{T}_i(t_i^0)=[\underline{t}_i^{f}, \overline{t}_i^{f}]}$ be the feasible range of terminal times under the state and input constraints of \CAV{i} computed at $t_i^0$. 
The formulation for computing $\underline{t}_i^{f}$ and $\overline{t}_i^{f}$ can be found in \cite{chalaki2020experimental}.
Then \CAV{i} solves the following time-optimal control problem to find the minimum exit time $t_i^{f} \in \mathcal{T}_i(t_i^0)$ that satisfies all state, input, and safety constraints
\label{prb:optimal_MZ}
\begin{align}
\begin{split}
\label{eq:tif_1}
&\minimize_{t_i^{f} \in \mathcal{T}_i(t_i^0)} \quad t_i^{f} 
\\
& \subjectto 
\\
& \quad \eqref{eq:uconstraint}, \eqref{eq:vconstraint}, \eqref{eq:lateral_constraint}, \eqref{eq:rearend_constraint},
\eqref{eq:optimalTrajectory},
\\
& \text{given:} 
\\
& \quad p_i (t_i^0) = p^0, v_i (t_i^0) = v_i^0, p_i (t_i^{f}) = p^{f}, u_i (t_i^{f}) = 0.
\end{split}
\end{align}
\end{problem}

\begin{algorithm}[t]
\caption{Numerical algorithm for solving Problem~\ref{prb:2}}
\label{alg:optimal_exit_time}
\begin{algorithmic}[1]
\For{$i = 1$ to $N(t)$}
\State Initialize $t_i^{f} \gets \underline{t}_i^{f}$
\Repeat
\State Compute the trajectory coefficients $\bbsym{\phi}_i$ using \eqref{eq:phi_i}
\State Evaluate the constraints \eqref{eq:uconstraint}, \eqref{eq:vconstraint}, \eqref{eq:lateral_constraint}, \eqref{eq:rearend_constraint}, \eqref{eq:optimalTrajectory}
\If{no constraint is violated}
    \State \textbf{return} $t_i^{f}$, $\bbsym{\phi}_i$
\Else
    \State $t_i^{f} \gets t_i^{f} + \epsilon$ 
\EndIf
\Until{$t_i^{f} \ge \overline{t}_i^{f}$}
\EndFor
\end{algorithmic}
\end{algorithm}\setlength{\textfloatsep}{6pt} 

The numerical procedure for solving Problem~\ref{prb:optimal_MZ} is summarized as follows and can also be found in \cite{chalaki2020experimental}. 
We initialize \( t_i^{f} = \underline{t}_i^{f} \), compute \( \boldsymbol{\phi}_i \) using \eqref{eq:phi_i}, and evaluate all state, control, and safety constraints. 
If the constraints are satisfied, the solution is accepted; otherwise, \( t_i^{f} \) is increased by a step size \( \epsilon \). This process is repeated until a feasible solution is obtained. 
Solving Problem~\ref{prb:optimal_MZ} yields the optimal exit time \( t_i^{f} \) and the corresponding optimal trajectory and control law in \eqref{eq:optimalTrajectory} for \CAV{i} over \( t \in [t_i^{0},\, t_i^{f}] \). 
The full procedure is given in Algorithm~\ref{alg:optimal_exit_time}. 
If the problem for \CAV{i} is infeasible, the algorithm returns the solution with minimal constraint violation.

\subsection{Replanning Mechanism}

Unlike our previous framework for CAV coordination \citep{malikopoulos2018decentralized}, 
where each CAV solves its optimization problem only upon entering the control zone, we enhance the framework by enabling the CAVs to re-solve the time-optimal trajectory planning 
problem at every discrete time step based on newly observed information. 
At each step, a CAV’s position and speed are updated and used as new initial conditions for 
solving the time-optimal trajectory planning problem (Problem~\ref{prb:2}). 
Incorporating such a replanning mechanism introduces feedback into the planning process 
and has been shown to improve closed-loop performance; see \citep{chalaki2021Reseq}.

At each time step, we need to determine the decision sequence for the CAVs.
Let $s = (s_1, s_2, \dots, s_{N(t)})$, where $s_i \in \{1,...,N(t)\}$ and $s_i \neq s_j$, $\forall\, i \neq j$, be the decision sequence for the CAVs, which determines the order at which the CAVs solve its trajectory planning problem. 
We compute the decision sequence based on the following rule:
\begin{equation}
\label{eq:dec_seq}
[\underline{t}_i^{f}, \overline{t}_i^{f}] \prec [\underline{t}_j^{f}, \overline{t}_j^{f}] 
\text{ then } s_i < s_j, 
\end{equation}
where $\prec$ denotes the comparison in lexicographic order.
The rule implies that CAVs closer to the control zone exit have higher priority. In cases where two CAVs have the same minimum possible exit time, the one with the smaller feasible space is assigned higher priority.

Given the generated decision sequence, the CAVs sequentially solve the time-optimal control problem. 
\begin{problem} 
(\textbf{Cooperative planning for all CAVs})
At any time step $t^{c}$, \CAV{i} $\in \AAA (t^{c})$, in a sequential manner determined by \eqref{eq:dec_seq}, solves the following time-optimal control problem
\label{prb:cc-replan}
\begin{align}
\begin{split}
&\minimize_{t_i^{f} \in \mathcal{T}_i(t^{c})} \quad t_i^{f} 
\\
& \subjectto 
\\
& \quad \eqref{eq:uconstraint}, \eqref{eq:vconstraint},  \eqref{eq:lateral_constraint}, \eqref{eq:rearend_constraint},
\eqref{eq:optimalTrajectory},
\\
& \text{given:} 
\\
& \quad p_i (t^{c}), v_i (t^{c}), p_i (t_i^{f}) = p^{f}, u_i (t_i^{f}) = 0. 
\end{split}
\end{align}
\end{problem}

At each replanning instance, we solve \(N(t)\) time-optimal control problems sequentially, which becomes increasingly expensive as the number of CAVs grows. 
Problem~\ref{prb:cc-replan}, together with the decision sequence \eqref{eq:dec_seq}, can be viewed as a parametric optimization problem whose optimal solutions depend on time-varying parameters, such as the initial conditions of the CAVs. 
To accelerate its solution, we learn a mapping from problem parameters to optimal solutions using supervised learning and offline-generated data. 
The trained model then provides approximate optimal solutions for real-time execution. 
In the next section, we develop a GNN-based learning framework to approximate the solutions of these time-optimal control problems.

\section{Learning to Warm-Start with Graph Neural Networks}
\label{sec:gnn}

In this section, we present a GNN-based learning framework for predicting the optimal terminal time $t_i^{f,*}$.
GNNs have been successfully employed in different applications for multi-agent coordination, such as learning binary solutions in multi-agent mixed-integer programming; see \citep{le2025multirobot}, multi-agent reinforcement learning; see \citep{goeckner2024graph}.
To this end, we use GINEConv, an extension of the graph isomorphism network (GIN) architecture that explicitly incorporates edge features; see \citep{xu2019how,hu2020pretraining}.
We first define the graph as follows.
\begin{definition}
\label{def:graph}(Graph representation) Let $\GGG = (\VVV, \EEE)$ be the graph representing the network of CAVs, in which $\VVV = \AAA(t)$ and $\EEE \subset \AAA(t) \times \AAA(t)$ are the node and edge sets, respectively.
We consider a directed edge set in which a directed edge $(i,j) \in \EEE$ is included if \CAV{i} is assigned higher priority than \CAV{j} in the decision sequence, \ie $s_i < s_j$, and the two CAVs are coupled by either a lateral safety constraint or a rear-end safety constraint. 
\end{definition}

\begin{definition}(Node and edge features)
Let the node feature vector for each \CAV{i} be defined as $\bbsym{\theta}_i = [p_i, v_i, \underline{t}_i^{f}, \overline{t}_i^{f}, \bbsym{o}_i^\top]^\top$, where $\bbsym{o}_i$ is a one-hot encoding of the lane index.
The edge feature vector $\bbsym{\theta}_{ij}$ for each edge $(i,j) \in \EEE$ is defined as a one-hot encoding of the type of the corresponding safety constraint.
Specifically, $\bbsym{\theta}_{ij} = [1,0]^\top$ if \CAV{i} and \CAV{j} are coupled by a lateral safety constraint, and $\bbsym{\theta}_{ij} = [0,1]^\top$ if they are coupled by a rear-end safety constraint.
\end{definition}

\subsection{GINEConv Networks}

GINEConv is a GNN architecture that extends graph isomorphism networks (GINs) by explicitly incorporating edge features into message passing.
This makes GINEConv suitable for graph-structured problems in which the relations between nodes carry important information, such as safety constraints between CAVs.
Let $\boldsymbol{h}_i^{(k)}$ denote the embedding of node $i$ at layer $k$ of the network, with $\boldsymbol{h}_i^{(0)} = \boldsymbol{\theta}_i$. 
At each layer $k$, GINEConv updates the node embedding by aggregating messages from incoming neighboring nodes while incorporating edge features as follows:
\begin{equation}
\begin{multlined}
\boldsymbol{h}_i^{(k)}
=
\mathrm{MLP}^{(k)}
\Bigg(
(1+\epsilon^{(k)}) \bbsym{h}_i^{(k-1)} \\
+ 
\sum_{j \in \NNN(i)}
\mathrm{ReLU}
\left(
\bbsym{h}_j^{(k-1)} + \bbsym{\theta}_{ji}
\right)
\Bigg),
\end{multlined}
\label{eq:gineconv}
\end{equation}
where $\mathcal{N}(i) = \{j \in \mathcal{V} \mid (j,i) \in \mathcal{E}\}$ denotes the set of incoming neighbors of node $i$, $\mathrm{MLP}^{(k)}(\cdot)$ is a multilayer perceptron at layer $k$, and $\epsilon^{(k)}$ is either a fixed scalar or a learnable parameter.
For full details on GINEConv, we refer the reader to \cite{hu2020pretraining}.

Let $\hat{t}_i^{f}$ be the prediction for the optimal terminal time ${t}_i^{f*}$.
Note that given the predicted optimal terminal time $\hat{t}_i^{f}$ from the GINEConv network, the corresponding energy-optimal trajectory for \CAV{i} can be found by solving \eqref{eq:phi_i}.
To ensure that the solution of \eqref{eq:phi_i} exists, we require the prediction $\hat{t}_i^{f*}$ must be inside the feasible range $[\underline{t}_i^{f}, \overline{t}_i^{f}]$ computed at each time step.
Thus, we consider the following computation for $\hat{t}_i^{f}$:   
\begin{equation}
\label{eq:tif_pred}
\hat{t}_i^{f} = \underline{t}_i^{f} + (\overline{t}_i^{f} - \underline{t}_i^{f})\; \sigma (\bbsym{h}_i^{K}),
\end{equation}
where we apply the sigmoid function to the output $\bbsym{h}_i^{K}$ of the GINEConv network.

\begin{algorithm}[t]
\caption{GNN-based accelerated numerical algorithm for solving Problem~\ref{prb:2}}
\label{alg:gnn-solver}
\begin{algorithmic}[1]
\State Make GNN prediction using \eqref{eq:tif_pred} to obtain $\{\hat{t}_i^{f}\}_{i \in \AAA(t)}$
\For{$i = 1$ to $N(t)$}
    \State Initialize $t_i^{f} \gets \hat{t}_i^{f}$
    \State Compute trajectory coefficients $\bbsym{\phi}_i$ using \eqref{eq:phi_i}
    \State Evaluate constraints \eqref{eq:uconstraint}, \eqref{eq:vconstraint}, \eqref{eq:lateral_constraint}, \eqref{eq:rearend_constraint}, \eqref{eq:optimalTrajectory}
    
    \If{no constraint is violated}
        \State Store $(t_i^{f}, \bbsym{\phi}_i)$ as the current feasible solution
        \While{$t_i^{f} - \epsilon \ge \underline{t}_i^{f}$}
            \State $t_i^{f} \gets t_i^{f} - \epsilon$
            \State Compute trajectory coefficients $\bbsym{\phi}_i$ using \eqref{eq:phi_i}
            \State Evaluate constraints \eqref{eq:uconstraint}, \eqref{eq:vconstraint}, \eqref{eq:lateral_constraint}, \eqref{eq:rearend_constraint}, \eqref{eq:optimalTrajectory}
            \If{no constraint is violated}
                \State Update the current feasible solution to $(t_i^{f}, \bbsym{\phi}_i)$
            \Else
                \State \textbf{break}
            \EndIf
        \EndWhile
        \State \textbf{return} the current feasible solution
    \Else
        \While{$t_i^{f} + \epsilon \le \overline{t}_i^{f}$}
            \State $t_i^{f} \gets t_i^{f} + \epsilon$
            \State Compute trajectory coefficients $\bbsym{\phi}_i$ using \eqref{eq:phi_i}
            \State Evaluate constraints \eqref{eq:uconstraint}, \eqref{eq:vconstraint}, \eqref{eq:lateral_constraint}, \eqref{eq:rearend_constraint}, \eqref{eq:optimalTrajectory}
            \If{no constraint is violated}
                \State \textbf{return} $t_i^{f}$, $\bbsym{\phi}_i$
            \EndIf
        \EndWhile
        \State Set $t_i^{f} \gets \overline{t}_i^{f}$
        \State Compute trajectory coefficients $\bbsym{\phi}_i$ using \eqref{eq:phi_i}
        \State \textbf{return} $t_i^{f}$, $\bbsym{\phi}_i$
    \EndIf
\EndFor
\end{algorithmic}
\end{algorithm}

\subsection{GNN-based Accelerated Numerical Algorithm}

As mentioned earlier, the trajectories generated by the GNN-based solver may be infeasible or suboptimal due to prediction errors relative to the true optimal solution.
Therefore, the GNN-based solver cannot be directly deployed for real-time control.
However, the predicted terminal time from the GNN can serve as a warm-start or initial guess to accelerate the optimization process.

The algorithm operates as follows.
For each \CAV{i}, we use the GNN prediction $\hat{t}_i^f$ as the initial solution.
If the initial solution is feasible, \ie satisfies all state, control, lateral safety, and rear-end safety constraints, the algorithm searches backward in increments of $\epsilon$ to find a smaller feasible terminal time until feasibility is lost or the lower bound is reached. 
If the initial solution is infeasible, the algorithm searches forward in increments of $\epsilon$ until it finds a feasible terminal time or reaches the upper bound.
If no feasible candidate is found, the algorithm returns the trajectory associated with the upper terminal-time bound.
The steps of the accelerated numerical solver are summarized in Algorithm~\ref{alg:gnn-solver}.

\section{Simulation Studies}
\label{sec:sim}

In this section, we demonstrate the control performance of the proposed framework through numerical simulations.
We created a simulation environment in \texttt{SUMO} interfacing with Python via TraCI; see \citep{lopez2018microscopic}.
In the simulation, we considered an intersection scenario with a control zone of length \SI{250}{m}.
The main program of our framework is implemented in Python, utilizing the PyTorch Geometric library for training and prediction of the GNN model. 
The parameters of the time-optimal control problem are chosen as:
$v_{\max} = \SI{20.0} {m/s}$, 
$v_{\min} = \SI{1.0} {m/s}$,
$a_{\max} = \SI{3.0} {m/s^2}$, 
$a_{\min} = \SI{-5.0} {m/s^2}$,
$\rho_{\rm l} = \SI{2.0}{s}$, 
$\rho_{\rm r} = \SI{1.0}{s}$,
$d_{\min} = \SI{7}{m}$.

\subsection{Training Details}

We adopt a DAgger-style imitation learning procedure to mitigate the distribution shift between offline training and deployment. 
We first train the GINEConv network using expert solutions generated by Algorithm~\ref{alg:optimal_exit_time}. 
The learned model is then deployed in simulation, and for each visited state, Algorithm~\ref{alg:optimal_exit_time} is queried to provide the corresponding expert label.
These newly labeled samples are aggregated with the original expert dataset and used to fine-tune the model, thereby exposing the network to the state distribution induced by its own predictions.
The GNN model consists of three GINEConv layers with $256$, $128$, and $64$ hidden units, respectively, followed
by a linear output layer for node-level regression. 
The network is trained by minimizing the Huber loss, which is less sensitive to outliers than MSE while retaining smooth quadratic behavior near small prediction errors. 
The final trained model achieves a training loss of $0.76$ and a validation loss of $0.10$.

\subsection{Simulation Results and Discussions}

\begin{figure}[tb]
\centering
\includegraphics[width=0.98\linewidth]{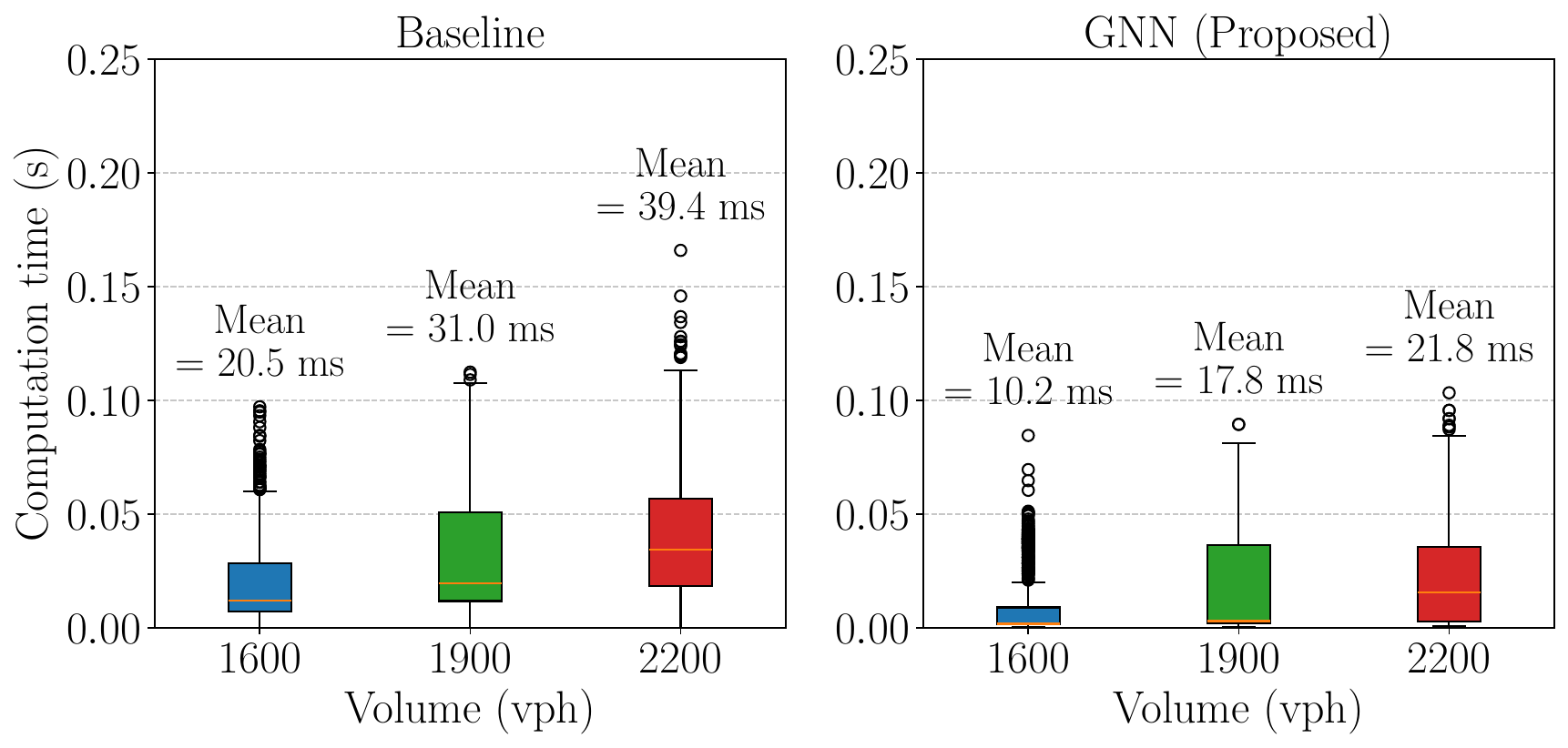}
\caption{Comparison of computation times across different traffic volumes for the baseline solver (Algorithm~\ref{alg:optimal_exit_time}) and the GNN-based solver (Algorithm~\ref{alg:gnn-solver}) under replanning at every time step.}
\label{fig:comp_times}
\end{figure}

To demonstrate the main advantage of the GNN-based solver, we compare its computation time with that of the baseline numerical solver (Algorithm~\ref{alg:optimal_exit_time}) under replanning at every time step, as shown in Fig.~\ref{fig:comp_times}. 
The results are reported for three traffic volumes: $1600$, $1900$, and $2200$ vehicles per hour. 
As shown in the figure, the GNN-based approach achieves an approximate $2.0\times$ speedup over the baseline numerical method.

\begin{table}[!tb]
\centering
\caption{Average travel time (and standard deviation), in seconds, for our proposed method (Algorithm~\ref{alg:gnn-solver} with replanning) and two baselines: Algorithm~\ref{alg:optimal_exit_time} with replanning (Baseline \#1) and without replanning (Baseline \#2).}
\label{tab:travel_time}
\begin{tabular}{
    >{\centering\arraybackslash}p{2.0cm}  %
    >{\centering\arraybackslash}p{1.6cm}  %
    >{\centering\arraybackslash}p{1.6cm}  %
    >{\centering\arraybackslash}p{1.6cm}  %
}
\toprule
\textbf{Methods} & \textbf{1600 veh/h} & \textbf{1900 veh/h} & \textbf{2200 veh/h} \\
\midrule
Proposed & $10.6$ ($1.66$) & $10.67$ ($1.62$) & $10.92$ ($1.66$) \\
\midrule
Baseline \#1 & $10.4$ ($1.35$) & $10.65$ ($1.38$) & $10.98$ ($1.32$)  \\
\midrule
Baseline \#2 & $11.44$ ($2.11$) & $12.15$ ($2.50$) & $18.71$ ($4.30$) \\
\bottomrule
\end{tabular}
\end{table}

\begin{figure}[tb!]
\centering
\begin{subfigure}[b]{0.24\textwidth}
\centering
\includegraphics[width=0.98\textwidth]{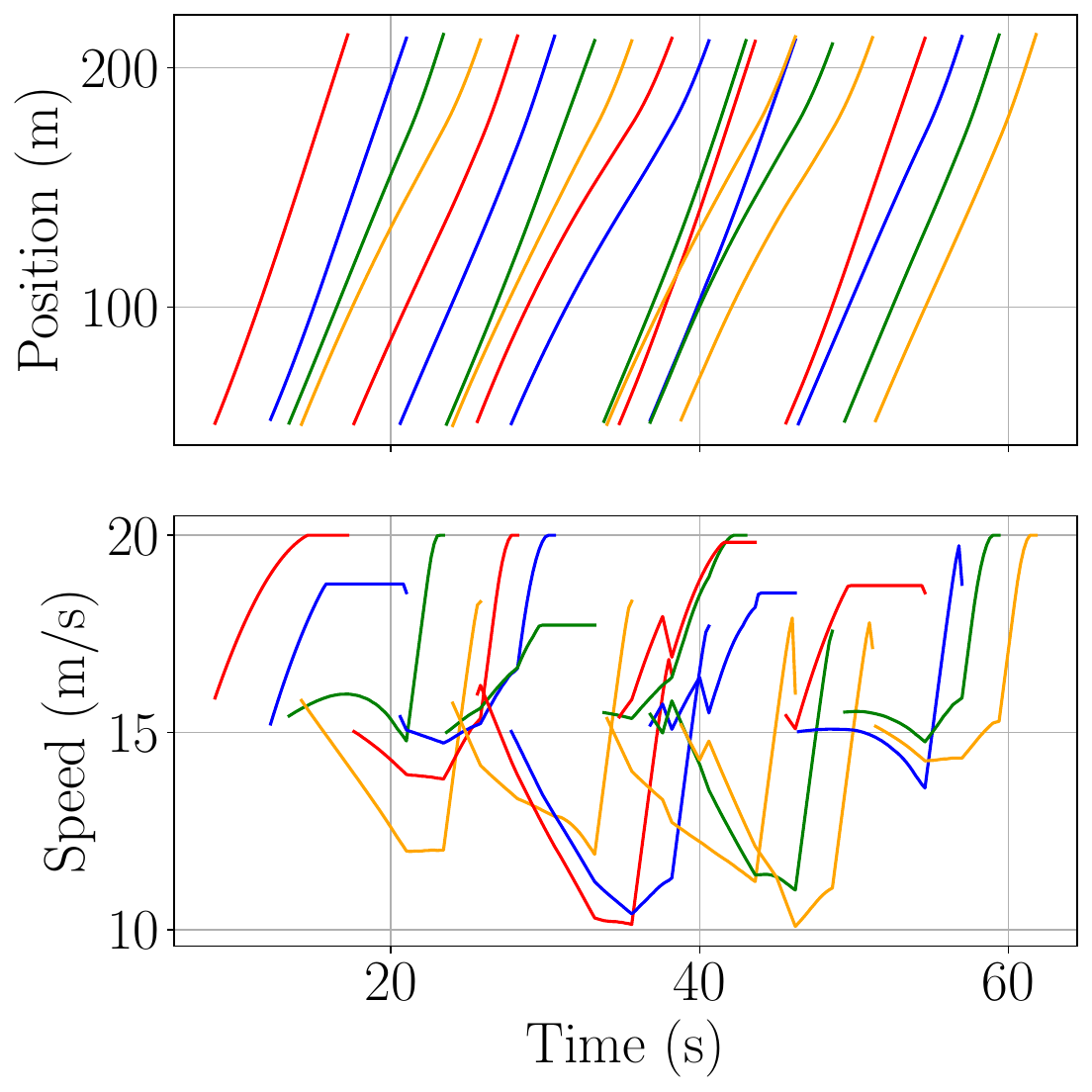}
\caption{GNN-based solver with replanning}
\label{fig:gnn_traj}
\end{subfigure}
\hfill
\begin{subfigure}[b]{0.24\textwidth}
\centering
\includegraphics[width=0.98\textwidth]{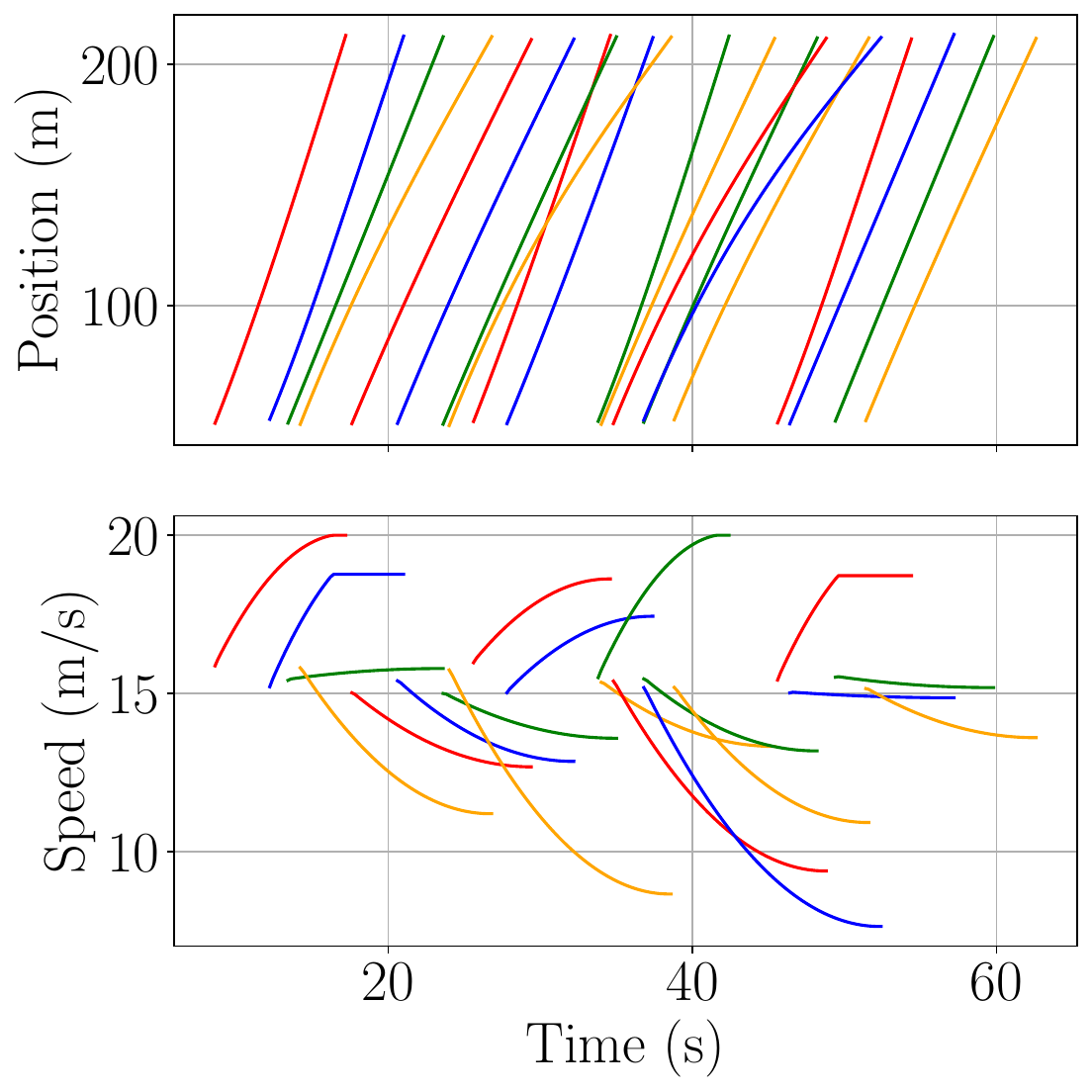}
\caption{Numerical solver without replanning}
\label{fig:noreplan_traj}
\end{subfigure}
\caption{Position and speed trajectories for $20$ vehicles in a simulation example using different methods. Different colors represent the trajectories of vehicles traveling on different lanes.}
\label{fig:traj}
\vspace{1mm}
\end{figure}

Next, we compare our proposed method, which performs replanning at every time step using Algorithm~\ref{alg:gnn-solver}, with two baseline methods in which the time- and energy-optimal control problem (Problem~\ref{prb:2}) is solved for each CAV at every time step (Baseline \#1) and only when it enters the control zone (Baseline \#2).
Table~\ref{tab:travel_time} shows the travel time comparison under traffic volumes of $1600$, $1900$, and $2200$ vehicles per hour. 
The results indicate that the proposed method yields average travel times comparable to Baseline~\#1, which also uses replanning, while consistently outperforming Baseline~\#2, which does not use replanning. 
In particular, relative to Baseline~\#2, the proposed method reduces the average travel time by approximately $7.3\%$, $12.2\%$, and $41.6\%$, respectively.
In addition, we show the position and speed trajectories of several CAVs in a particular simulation in Fig~\ref{fig:traj}.
The speed profiles show that, without replanning, the CAVs can accelerate or decelerate throughout the control zone, which is consistent with the solution of Problem~\eqref{prb:2}. 
In contrast, when replanning is enabled, the CAVs can improve their trajectories whenever a more efficient yet still safe solution is available.

In addition to the results reported in this paper, we include simulation videos and code implementation at \url{https://sites.google.com/seas.upenn.edu/cav-gnn}.

\section{Conclusion}
\label{sec:cls}

In this paper, we developed a \emph{learning-to-warm-start} frameworkthat accelerates the solution of the time-optimal trajectory planning problem for CAVs in an unsignalized intersection. 
The framework leverages GINEConv, a variant of GNNs, to learn the optimal solutions to a multi-agent time-optimal trajectory planning problem and to provide high-quality warm-start predictions for real-time implementation. 
Unlike previous approaches, the proposed method enables real-time replanning by solving the time-optimal trajectory planning problem at every discrete time step. 
Future work will extend this framework to mixed-traffic environments with human-driven vehicles.

\bibliography{references,references_IDS}

\end{document}